Phase separation and stripe patterns in K$_{0.8}$Fe$_{1.6+x}$Se$_2$ superconductors

Z. W. Wang, Z. Wang, Y. J. Song, C. Ma, H. L. Shi, Z. Chen, H. F. Tian, H. X. Yang, G. F. Chen and J. Q. Li*

Beijing National Laboratory for Condensed Matter Physics, Institute of Physics, Chinese Academy of Sciences, Beijing 100190, P. R. China

Structural investigations on the K$_{0.8}$Fe$_{1.6+x}$Se$_2$ superconducting materials have revealed remarkable micro-stripes arising evidently from the phase separation. Two coexisted structural phases can be characterized by modulations of $q_1 = 1/5[a^*+3b^*]$, the antiferromagnetic phase K$_{0.8}$Fe$_{1.6}$Se$_2$, and $q_2 = 1/2[a^*+b^*]$, the superconducting phase K$_{0.75}$Fe$_2$Se$_2$, respectively. These stripe patterns likely result from the anisotropic assembly of superconducting particles along the [110] and [1-10] direction. In addition to the notable stripe structures, a nano-scale phase separation also appears in present superconducting system as clearly observed by high-resolution transmission electron microscopy. Certain notable experimental data obtained in this heterogenous system can be quantitatively explained by the percolation scenario.

PACS number: 74.62. Bf, 74.60.-w, 74.62.-c, 61.14.Lj, 61.16.Bg

Following the discovery of superconductivity in LaFeAsO$_{0.89}$F$_{0.11}$ with T$_c$ = 26K, a number of Fe-based superconductors with different structures have been obtained, such as the '111', the '122' and the '1111'-type systems [1-4]. Recently, in addition to the FeAs-layer materials, superconductivity has been observed in the Fe-deficient FeSe-layered structure above 30K [5]. Actually, this superconducting system has a defect-driven 122 structure (so called 122* phase) with nominal composition of A$_y$Fe$_z$Se$_2$ (A = K, Rb, Tl and Cs) [6-9]. The 122* materials in general exhibit complex microstructure and magnetic properties resulting from the Fe-deficiency and vacancy ordering [10-13]. Indeed, numerous experimental and theoretical studies showed a visible structural inhomogeneity in superconducting samples [11, 14-17], these phenomena have certain similarities with the results observed in the high-T$_c$ cuprate superconductors and colossal magnetoresistance (CMR) manganese [18-20]. It has been also noted that the transition between the antiferromagnetic parent phase K$_{0.8}$Fe$_{1.6}$Se$_2$ and the superconductors, along with the increase of Fe-content, occurs in association with a mixed-phase process. Though a number of investigations of structural and physical properties has been reported in the 122* system, the structural phase and phase separation scenario has not been well identified, therefore certain fundamental issues in correlation with the superconducting mechanism are still in debate [12, 17]. In this paper, we will report on the microstructural properties of superconducting materials in the defect-driven 122* system. Our results demonstrate that a superstructure phase with a modulation of q$_2$ = 1/2[a*+b*] can be considered as superconducting phase, which in general coexists with the known antiferromagnetic K$_{0.8}$Fe$_{1.6}$Se$_2$ parent phase and results in a micron-scale stripe patterns and an irregular nano-scale domains in the phase-separated state. The joint effects of two scales phase separations could play a critical role for understanding the physical properties observed in present system.

A series of well-characterized single-crystal samples were used in the present study. The single crystals were grown by the Bridgeman method [6], and polycrystalline samples were prepared by a two-step method [22]. In particular, materials with nominal composition of $K_{0.8}Fe_{1.6+x}Se_2$ have been extensively investigated to understand the alteration of microstructure and physical properties in the phase-separated state. Specimens for TEM observations were prepared by peeling off a very thin sheet, several tens of microns thick, from the single crystal and then crushing it in an agate mortar. High-resolution TEM observations and microstructure analysis were performed on a Tecnai-F20 transmission electron microscope.

The clear view of the phase-separation nature and related structural characterizations was performed on $K_{0.8}Fe_{1.6+x}Se_2$ single crystals by using scanning electron microscopy (SEM) and scanning transmission electron microscopy (STEM). The most remarkable microstructure feature revealed in our observations is the appearance of stripe-like patterns in all superconducting samples as clearly shown in Fig. 1(a-d). It is clearly recognizable that the concentration of these dark-contrast stripes increases progressively with the increase of Fe content. In Fig. 2, we show the experimental data obtained from STEM associated with the chemical composition analysis for $K_{0.8}Fe_{1.7}Se_2$ sample. These experimental results demonstrate that, in general, the stripes go along the [110] and [1-10] directions and consist of a number of assembled grains. Chemical composition analyses suggest that these grains have a composition of about $K_{0.75}Fe_2Se_2$, which can be also estimated by the average period of the superconducting samples. Based on the SADPs in Fig. 2(b) and (c), we can conclude that the areas with light contrast are mainly governed by the antiferromagnetic parent phase and the superstructure with modulation of $q_2 = 1/2[a^*+b^*]$ is more concentrated at the stripe areas. It is believed that the appearance of superconductivity in present system essentially in correlation with the emergence of these micro-stripes. In additional to the stripe pattern, our careful observations show that the phase separation on the nano-scale also appears in present system as illustrated in the SEM image of Fig. 3(a). This issue will be further analyzed in following context by using the high-resolution TEM.

The presence of chemical and structural defects in the 122* superconducting materials is a critical issue for understanding the fundamental properties of this notable superconducting system. For instance, the Fe-vacancy ordering and structural inhomogeneities in this layered system actually play a critical role in its magnetic and superconducting behavior [13, 23]. It has also been well demonstrated that a superstructure phase of $K_{0.8}Fe_{1.6}Se_2$, the so-called 245 phase, is a fundamentally important one, which exhibits a notable antiferromagnetic state ($T_N \sim 560K$) in association with a well-defined Fe-vacancy order [10]. This Fe-vacancy order appears in the FeSe-layers, characterized by a modulation of $q_1 = 1/5\,[a^*+3b^*]$. Importantly, in addition to the micro-stripe patterns, our TEM investigations of the well-characterized superconducting samples in general show nano-scale irregular domains arising from the coexistence of the 245 phase with the $q_2 = 1/2[a^*+b^*]$ phase. Figure 3(b) and (c) show two typical diffraction patterns taken from a superconducting sample with nominal composition of $K_{0.8}Fe_{1.75}Se_2$, illustrating the coexistence of the two superstructures ($q_1$ and $q_2$) in one crystal. Note that the $q_2$ modulation, in contrast with $q_1$, often shows a short coherence feature along the $c^*$-axis direction. Our high-resolution TEM observations and theoretical simulation demonstrate that increased Fe concentration not only fills up a fraction of the Fe-vacancies but also results in a stable structure with K-ordering [17], and the experimental observations on the $q_2$-phase suggest a composition of around $K_{0.75}Fe_2Se_2$ in which the K ions show a double-spaced order within the a-b plane [17]. Figure 3(d) and (e) show two structural models schematically illustrating the double-spaced order in the K layer for $K_{0.75}Fe_2Se_2$ in comparison with the 245 phase with $q_1$-order in the FeSe-layer. We have come to believe that this $q_2$-phase is responsible for the appearance of superconductivity in the present materials.

We actually have prepared a series of polycrystalline and single-crystal samples of $K_{0.8}Fe_{1.6+x}Se_2$ with $0 \leq x \leq 0.4$. X-ray diffraction measurements and TEM observations reveal that the main diffraction peaks from materials with $0 \leq x \leq 0.20$ can be indexed as the known layered structure with lattice parameters of $a = b = 3.913$ Å, and $c = 14.03$ Å. However, samples with $x > 0.2$ in general are not single-phase, the existence of Fe grains as a typical impurity phase can be recognized in the X-ray diffraction patterns and also identified in the EDAX analysis. This fact demonstrates that Fe-vacancies in $K_{0.8}Fe_{1.6+x}Se_2$ could not be efficiently filled by increasing Fe content in the range of x >

0.2, in contrast with what has been observed for x ≤ 0.2. Figure 4(a) shows the X-ray diffraction data for $K_{0.8}Fe_{1.6+x}Se_2$, it is clearly recognizable that a series of additional diffraction peaks become visible following with the appearance of superconductivity as indicated by arrows, this structural phase not only yields a micro-stripe pattern but also leads a remarkable nano-phase separation. Its lattice parameter can be calculated to be extended ~1% along c-axis compared to the parent phase, i.e. $c$ = 14.19 Å [16]. Figure 4(b) shows the average ICP (Inductive Coupled Plasma-mass Spectrometer) data obtained from a few sets of single crystals of $K_{0.8}Fe_{1.6+x}Se_2$, and each set has been grown under slightly different conditions. Note that the Fe content could be rather different from the nominal composition for the samples with relatively high Fe content. These results are fundamentally in agreement with the X-ray diffraction data discussed above.

Figure 5(a-c) show the temperature dependence of the in-plane resistivity for a series of selected $K_{0.8}Fe_{1.6+x}Se_2$ single crystals. Noted that the resistivity behavior obviously depends on the Fe concentration, and superconductivity is observed in samples with x > 0.06. The most striking feature observed in our experiments is the appearance of two critical Fe-contents at x = 0.05 and 0.13 respectively. When x is less than 0.05, the resistivity curves for all samples increase continuously with decreasing temperature and show semiconducting behavior like that observed for the antiferromagnetic parent phase (x = 0). When x is in the range of 0.06 ≤ x ≤ 0.13, as temperature decreases, the resistivity of the crystals first increases, then peaks, and decreases below 200 K. This kind of hump-like behavior is commonly observed in the ρ(T) curve for the K-Fe-Se superconductors [17, 24]. Superconducting transition in this range occurs at about $T_c$ = 31 K, where the $T_c$ is chosen as the onset temperature of the downturn in resistivity, as labeled by the arrow in Fig. 5(b). On the other hand, when the Fe content increases above x ≥ 0.125, all samples show multi-superconducting transitions, this phenomenon has also been noted in previous literature [6]. In most samples, two visible superconducting transitions can be clearly recognized: one appears at the temperature about 31K and the other appears at $T_c$ = 43 K, as indicated in Fig. 5(c).

Figure 5(d) shows the temperature dependence of DC magnetic susceptibility – both zero-field-cooled (ZFC) and field-cooled (FC) for samples of $K_{0.8}Fe_{1.7}Se_2$ (x = 0.1) and $K_{0.8}Fe_{1.75}Se_2$ (x = 0.15) measured under applied external magnetic field of 10 Oe. As mentioned above, $K_{0.8}Fe_{1.7}Se_2$ has a sharp transition at 31K and, in contrast, $K_{0.8}Fe_{1.75}Se_2$ exhibits two broad superconducting transitions. It is also noted that although the $K_{0.8}Fe_{1.75}Se_2$ sample shows visible drop in resistivity at about 43 K, no clear diamagnetic feature was observed in our susceptibility data. This fact demonstrates that the main superconducting phase in the 122* layered system has a critical transition around $T_c \approx 31$ K.

We now proceed to discuss the alteration of microstructure and the nature of the nano-scale phase-separated state in $K_{0.8}Fe_{1.6+x}Se_2$ materials. Actually, our TEM observations of $K_{0.8}Fe_{1.6+x}Se_2$ offer a rich variety of structural phenomena, unrivaled by those of any other Fe-based superconducting system. The most striking structural phenomenon is the notable structural inhomogeneity in areas as small as a few nanometers in size; the insulator-to-superconductor transformations with increasing Fe content in $K_{0.8}Fe_{1.6+x}Se_2$ can be fundamentally interpreted as percolative transport through the superconducting domains.

In order to better view the phase-separated patterns for the superconducting materials with different Fe-contents, we have taken a variety of dark-field TEM images on a few well-characterized superconducting samples. The $q_1$-modulation is known to be an essential character of the parent phase of $K_{0.8}Fe_{1.6}Se_2$ (x = 0), thus its dark-field image could directly reveal the Fe-ordered domains embedded in the superconducting crystals. Despite the Fe-ordered component, the Fe-vacancies in other portions of a crystal have partially filled and the Fe-ordered state has been locally destroyed. Taking into account the remarkable superconductivity in these samples, we can conclude that these portions are mainly governed by the superconducting phase. In Fig. 6(a) and (b), we show two typical dark-field images by using $q_1$-superstructure spots for the superconducting samples with x = 0.08 and 0.15, respectively, exhibiting the complex contrast in phase-separated states which can be explained directly by a two-phase picture. In order to better view the superconducting network in the phase-separated state, we have artificially colored the dark-field TEM images so that the Fe-vacancy order regions are faded and, on the other hand, superconducting networks are shown in relatively bright contrast in resultant images. It is clear that the nanometer-scale phase-separated regions with fine speckled or striped contrast can be commonly observed for superconducting samples with $0.06 \leq x \leq 0.12$. On the other hand, it is known that samples with $x > 0.12$ often exhibit multiple transitions as mentioned above. Microstructure analysis shows that the spherical droplets with relatively large diameters in the phase-separated state can be frequently observed. In order to directly display the atomic scale structural features in the phase-separated state, Fig. 6(c) shows a high-resolution TEM image illustrating the phase-separated pattern within the a-b plane for an $x = 0.08$ sample, in which nano-domains of Fe-ordered and -disordered states are plainly visible. Moreover, careful investigations of the phase separations in the 122*-superconducting system often reveal a visible layered tendency, as exhibited in the high-resolution TEM image of Fig. 6(d) and discussed partially in our previous papers [11].

Phase separation as a significant phenomenon in strongly correlated systems has been a concern in a number of phenomenological approaches that can be used to predict the shapes of the clusters and even their electric transport properties [21, 25]. In the $K_{0.8}Fe_{1.6+x}Se_2$ superconducting system, it is noted micro-stripe patterns are likely in correlation with spinodal decomposition, however, the sub-structure of the stripe pattern show visible chessboard features from assembly of superconducting particles. The nano-scale phase separation is believed to arise from the competition between the local lattice stress and Coulomb repulsion among the Fe-vacancies, and the resultant phase separation clusters in general are not static but fluctuating [17]. Our experimental observations reveal that the insulator-to-superconductor transition in $K_{0.8}Fe_{1.6+x}Se_2$ can be fundamentally interpreted as percolative transport through the superconducting domains. Based on the three-dimensional percolation model, the threshold is estimated to be in the range of 10–25% for superconducting fraction in percentage [25]. In our case, the threshold for superconductivity can be simply estimated based on the decrease of Fe-vacancies: i.e. 0.06/0.4 = 15% for the x = 0.06 sample, this fact suggests that $K_{0.8}Fe_{1.6+x}Se_2$ shows a superconductivity threshold at about 15% in agreement with the percolation model. In Fig. 6(e) and (f), we present two simplified structural models illustrating the alterations of nano-domain structures and the shapes of the clusters in percolative networks. Note that the phase separation patterns change in an obvious way with the increase of Fe content in the superconducting sample.

In summary, microstructure investigations of $K_{0.8}Fe_{1.6+x}Se_2$ superconducting system reveal a joint phase-separation scenario occurring on both the micron- and nano - scales in this superconducting system. The phase-separated pattern on micron scale often adopts notable stripe configurations along the [110] and [1-10] direction, on the other hand, the nano-scale phase separation yields irregular structural domains. Two coexisted structural phases can be characterized by the modulations of $q_1 = 1/5[a*+3b*]$, the antiferromagnetic $K_{0.8}Fe_{1.6}Se_2$ phase, and $q_2 = 1/2[a*+b*]$, the superconducting phase, respectively. The $q_2$-phase with composition of $K_{0.75}Fe_2Se_2$ is considered the superconducting phase playing a critical role for understanding of the essential physical properties in present system. Careful analysis suggests that the phase-separated materials can be written as $K_{0.8}Fe_{1.6+x}Se_2$ which shows up the superconducting $T_c \approx$ 31K for x ranging from 0.05 to 0.2. Though certain notable properties can be fundamentally explained by the percolation scenario, in order to better understand the complex structural and physical properties in this Fe-deficient system, a new mechanics for the phase separations should be developed in further study.

**Acknowledgements** This work is supported by the National Science Foundation of China under contracts No. 10874227, No.90922001, No. 10874214, No. 11074292 and No. 10904166, the Knowledge Innovation Project of the Chinese Academy of Sciences, and the 973 and 863 projects of the Ministry of Science and Technology of China.

Figure captions

Fig. 1 SEM images reveal the phase-separated state in $K_{0.8}Fe_{1.6+x}Se_2$ superconducting samples with (a) x = 0, (b) 0.06, (c) 0.15 and (d) 0.2, respectively, demonstrating the presence of stripe-like patterns along [110] and [1-10] direction. It is recognizable that stripe density increases visibly with the increase of Fe concentration.

Fig. 2 (a) STEM image of $K_{0.8}Fe_{1.7}Se_2$ along the c-axis direction shows the stripe pattern with darker contrast doing along the [110] and [1-10] direction. Selected area diffraction patterns taken from a light area (b) with modulated vector of $q_1=1/5[a^*+3b^*]$ and the stripe area (c) with modulation of $q_2=1/2[a^*+b^*]$. Elemental mapping of the rectangular area in Fig. 2(a) for (d) K element and (f) Fe element respectively, illustrating the higher Fe concentration for $q_2$ -phase.

Fig. 3 (a) SEM image shows the complex contrast from both micron- and nano-scale phase separations in $K_{0.8}Fe_{1.75}Se_2$. (b) and (c) SADPs from a superconducting sample taken along the [001] and [1-30] zone-axis directions, illustrating the coexistence of $q_1$ and $q_2$ modulations in the phase-separated state. (d) and (e), structural models exhibiting the K-vacancy order for the $q_2$-phase and Fe- vacancy order for the $q_1$-phase.

Fig. 4 (a) XRD patterns of $K_{0.8}Fe_{1.6+x}Se_2$ showing the coexistence of two structural phases in superconducting materials. (b) ICP data obtained from $K_{0.8}Fe_{1.6+x}Se_2$ single crystals.

Fig. 5 (a) Temperature dependence of electric resistivity for semiconducting $K_{0.8}Fe_{1.6+x}Se_2$ samples with x = 0 and 0.05. (b) Temperature dependence of electric resistivity for superconducting samples ($0.06 \leq x \leq 0.13$) showing $T_c$ at around 31K. (c) Temperature dependence of electric resistivity for superconducting samples ($x \geq 0.13$) with multi- superconductivity transitions. (d) Temperature dependence of ZFC and FC magnetic susceptibility of $K_{0.8}Fe_{1.6+x}Se_2$ single crystals with x = 0.1 and 0.15.

Fig. 6 TEM images show the phase-separated pattern in the $K_{0.8}Fe_{1.6+x}Se_2$ superconductors in (a) x = 0.08 and (b) x = 0.15. The fine speckled or striped contrasts can be recognized in the x = 0.08 superconductor. (c) High resolution TEM image shows nano-scale phase separated domains in the x = 0.08 sample. Note that in all superconducting samples phase separation exhibits a visible layering tendency, as illustrated in Fig.6 (d), the alternative appearance of Fe-ordered sheet (OS) and disordered sheet (DOS) are demonstrated. Qualitative representation of the microscopic phase-separated state in (e) x = 0.08 and (f) x = 0.15. The spherical droplets with a diameter of ~100nm appear commonly in samples with relatively high Fe content.

Fig.1

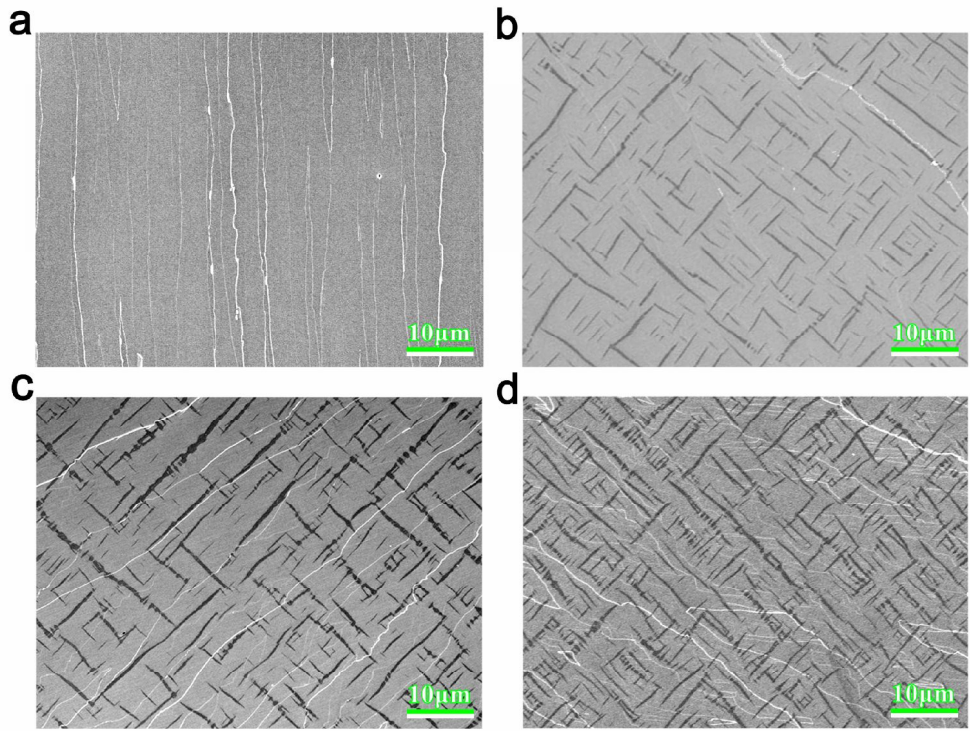

Fig.2

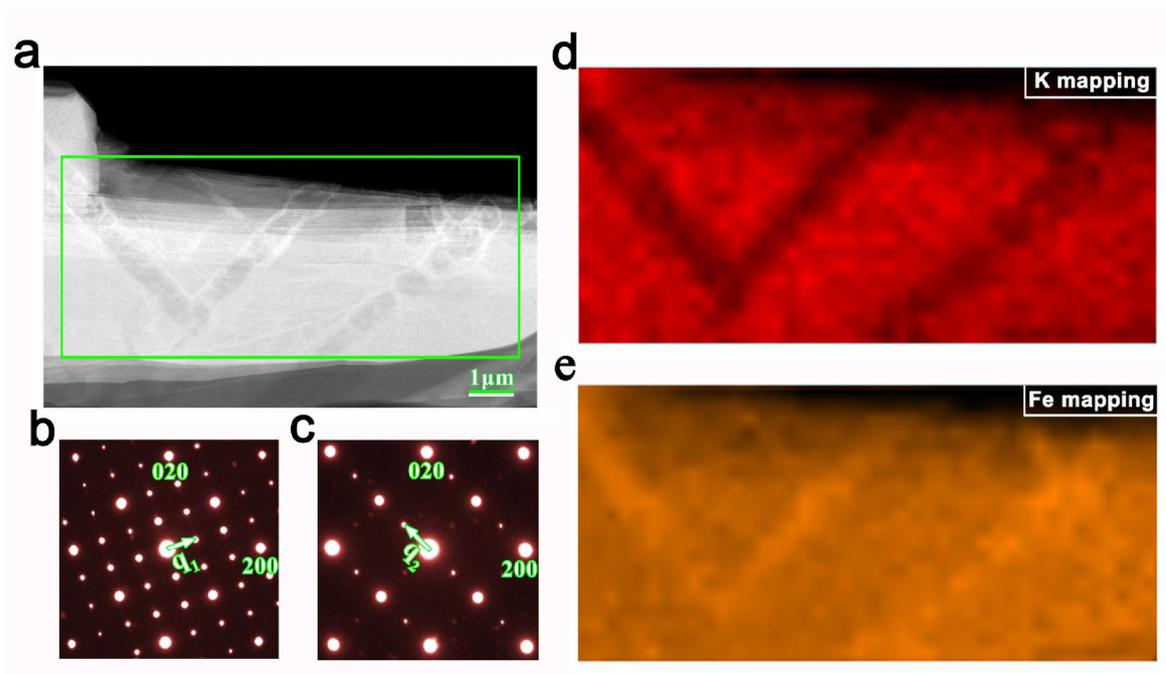

Fig. 3

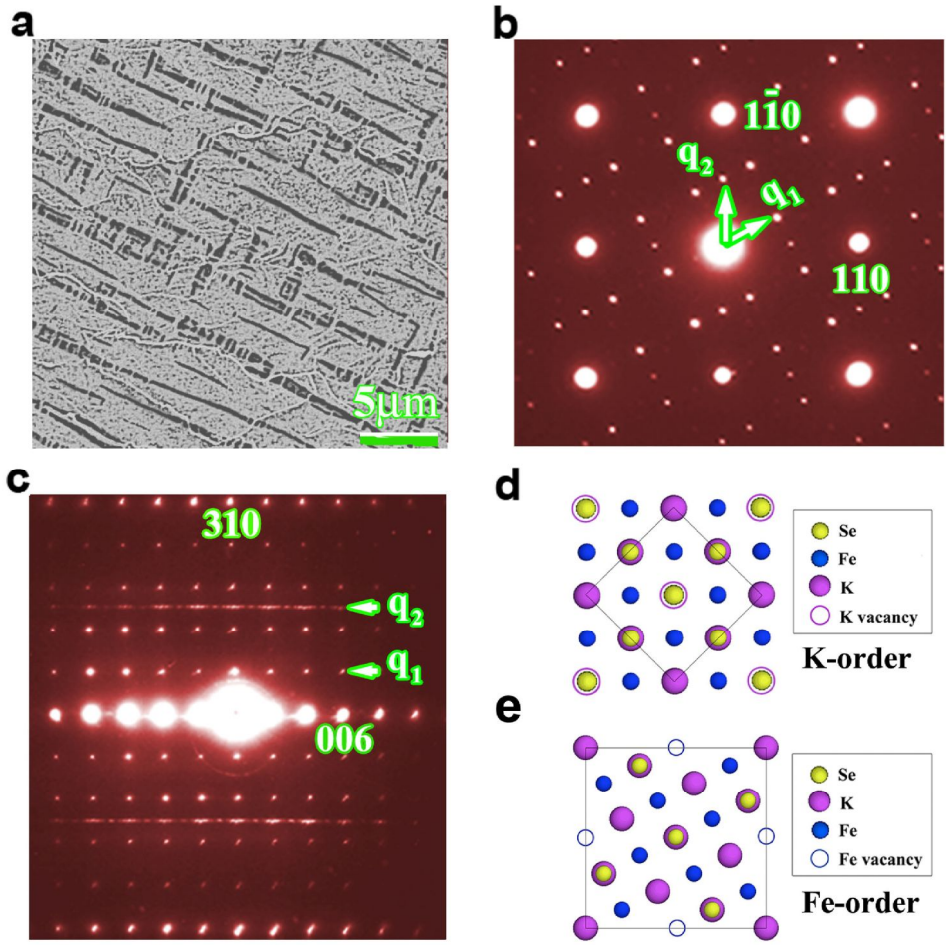

Fig.4

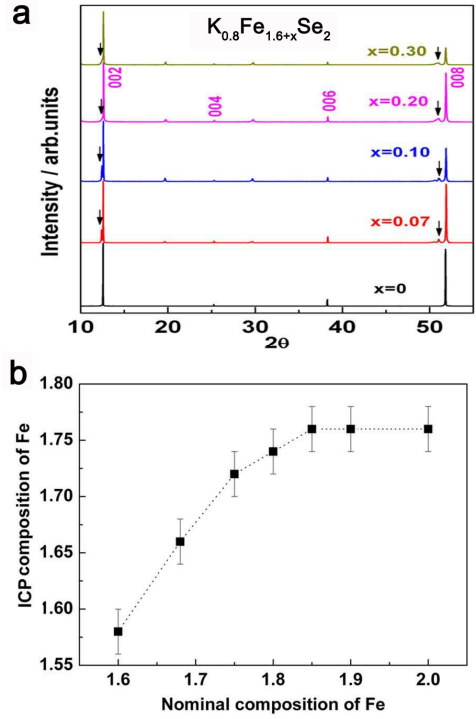

Fig. 5

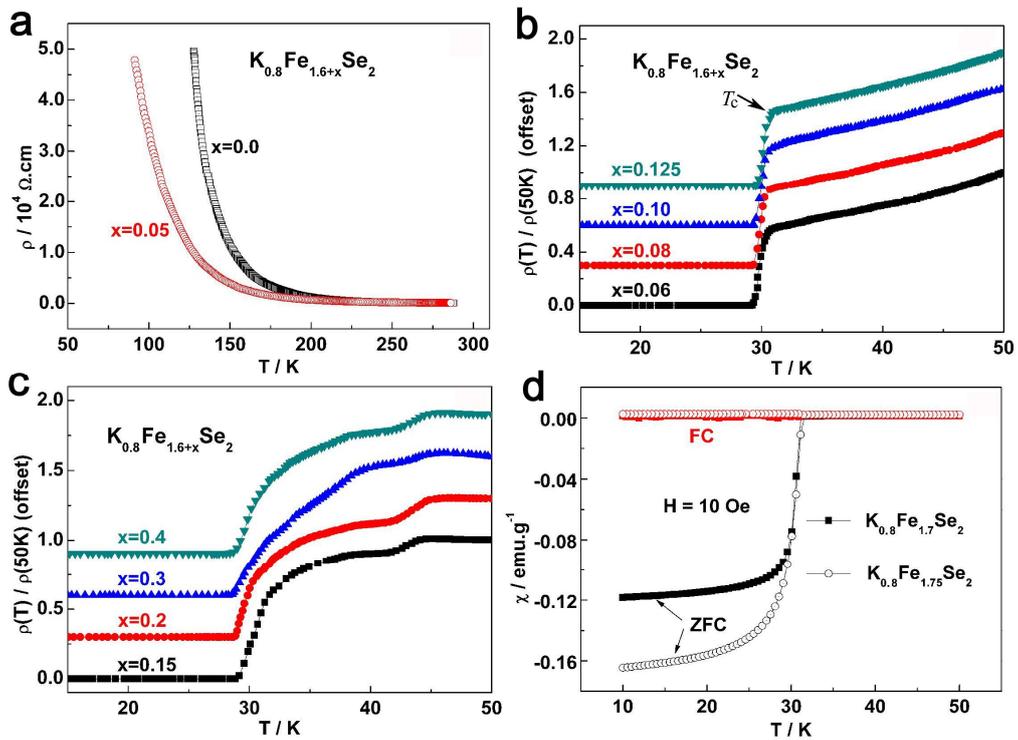

Fig.6

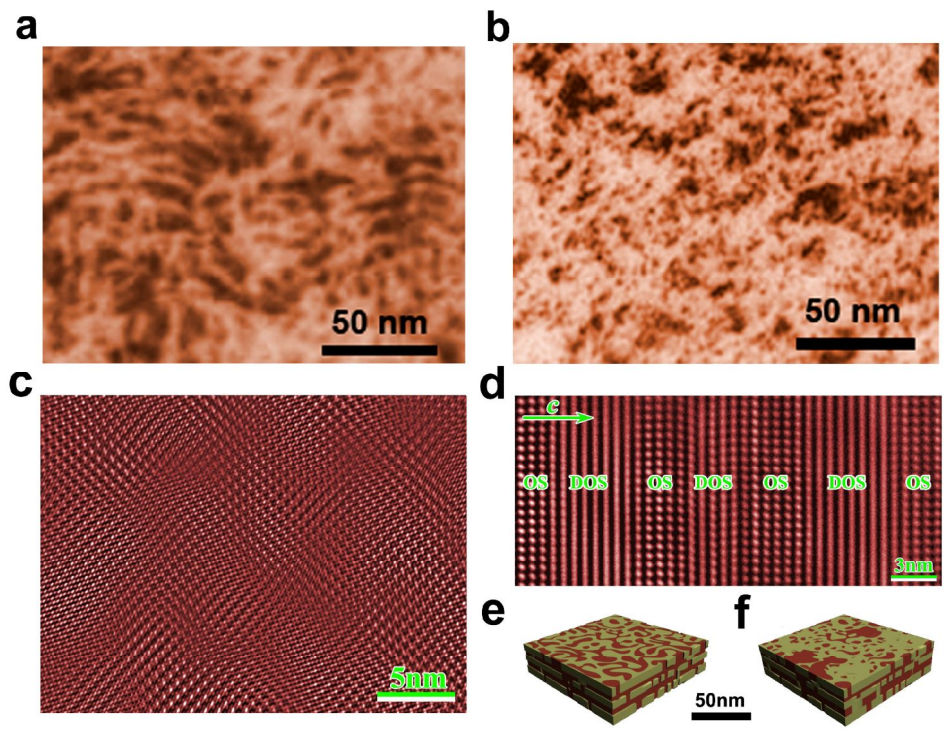